\documentclass[conference]{IEEEtran}
\IEEEoverridecommandlockouts
\usepackage{cite}
\usepackage{amsmath,amssymb,amsfonts}
\usepackage{algorithm}
\usepackage{algorithmic}
\usepackage{graphicx}
\usepackage{arydshln}
\usepackage{textcomp}
\usepackage{xcolor}
\usepackage{booktabs}
\usepackage{hyperref}
\usepackage{multirow}
\def\BibTeX{{\rm B\kern-.05em{\sc i\kern-.025em b}\kern-.08em
    T\kern-.1667em\lower.7ex\hbox{E}\kern-.125emX}}
\begin{document}

\title{Concealed Electronic Countermeasures of Radar Signal with Adversarial Examples}

\author{
\IEEEauthorblockN{Ruinan Ma}
\IEEEauthorblockA{\textit{School of Cyberspace} \\
                  \textit{Science and Technology} \\
                  \textit{Beijing Institute of Technology}\\
                  Beijing, China \\
                  ruinan@bit.edu.cn}
\and
\IEEEauthorblockN{Canjie Zhu}
\IEEEauthorblockA{\textit{School of Information and} \\
                  \textit{Electronics} \\
                  \textit{Beijing Institute of Technology}\\
                  Beijing, China \\
                  canjie@bit.edu.cn}
\and
\IEEEauthorblockN{Mingfeng Lu\textsuperscript{*}\thanks{* Corresponding author.}}
\IEEEauthorblockA{\textit{School of Information and} \\
                  \textit{Electronics} \\
                  \textit{Beijing Institute of Technology}\\
                  Beijing, China \\
                  lumingfeng@bit.edu.cn}
\and
\IEEEauthorblockN{Yunjie Li}
\IEEEauthorblockA{\textit{School of Information and} \\
                  \textit{Electronics} \\
                  \textit{Beijing Institute of Technology}\\
                  Beijing, China \\
                  liyunjie@bit.edu.cn}
\and
\IEEEauthorblockN{Yu-an Tan}
\IEEEauthorblockA{\textit{School of Cyberspace} \\
                  \textit{Science and Technology} \\
                  \textit{Beijing Institute of Technology}\\
                  Beijing, China \\
                  tan2008@bit.edu.cn}
\and
\IEEEauthorblockN{Ruibin Zhang}
\IEEEauthorblockA{\textit{School of Information and} \\
                  \textit{Electronics} \\
                  \textit{Beijing Institute of Technology}\\
                  Beijing, China \\
                  zhangruibin@bit.edu.cn}
\and
\IEEEauthorblockN{Ran Tao}
\IEEEauthorblockA{\textit{School of Information and} \\
                  \textit{Electronics} \\
                  \textit{Beijing Institute of Technology}\\
                  Beijing, China \\
                  rantao@bit.edu.cn}
\and
\IEEEauthorblockN{\textcolor{white}{Ruinan Ma}}
\IEEEauthorblockA{\textit{\textcolor{white}{School of Cyberspace}} \\
                  \textit{\textcolor{white}{Electronics}} \\
                  \textit{\textcolor{white}{Beijing Institute of Technology}}\\
                  \textcolor{white}{Beijing, China} \\
                  \textcolor{white}{ruinan@bit.edu.cn}}
}

\maketitle

\begin{abstract}
Electronic countermeasures involving radar signals are an important aspect of modern warfare. Traditional electronic countermeasures techniques typically add large-scale interference signals to ensure interference effects, which can lead to attacks being too obvious. In recent years, AI-based attack methods have emerged that can effectively solve this problem, but the attack scenarios are currently limited to time domain radar signal classification. In this paper, we focus on the time-frequency images classification scenario of radar signals. We first propose an attack pipeline under the time-frequency images scenario and DITIMI-FGSM attack algorithm with high transferability. Then, we propose STFT-based time domain signal attack(STDS) algorithm to solve the problem of non-invertibility in time-frequency analysis, thus obtaining the time-domain representation of the interference signal. A large number of experiments show that our attack pipeline is feasible and the proposed attack method has a high success rate.
\end{abstract}

\begin{IEEEkeywords}
electronic countermeasures, radar signal classification, time-frequency analysis, adversarial attack
\end{IEEEkeywords}

\section{Introduction}
Radar is widely deployed on the current battlefield. The classification technology of the radar signal has become the dominant key technology\cite{radar1, radar2}. On the one hand, recognizing the type of radar signals rapidly and accurately can obtain battlefield information and provide support for subsequent decision. On the other hand, adding interference to the original radar signal makes the receiver make a wrong judgment after receiving it, which is an effective attack method.\par

Deep learning represented by deep neural networks(DNN) has progressed rapidly in the last few years\cite{convnext, vit, resnet, mobilenet}. Converting one-dimensional time domain radar signals into two-dimensional time-frequency images through time-frequency analysis and then automatically extracts features from time-frequency images of different radar signals by trained DNN has become a conventional way for radar signal classification\cite{cr1, c1}. Scholars always focus on combining novel components with current networks for higher classification accuracy\cite{c1} or designing lightweight models suitable for deployment on mobile devices\cite{cr4}. Meanwhile, some scholars have begun to launch adversarial attacks in the time domain radar signal classification scenario based on DNN\cite{x1, x2}. By generating adversarial examples through attack algorithms, which are signals with imperceptible perturbations added, the model can make incorrect judgments. As far as we know, current attack methods of this kind are limited to the time-domain radar signal classification scenario and have not yet appeared in the attack work of the time-frequency analysis scenario, which has long been an important means of radar signal classification.\par

In this paper, we take the perspective of the radar signal transmitter and focus on attacks on the receptor using DNN for radar signal classification in the time-frequency analysis scenario. We refer to this as a new paradigm of electronic countermeasures, which we call concealed electronic countermeasures. We first propose a specific pipeline for launching attacks on the time-frequency images of radar signals. Then, in order to obtain the time domain representation of the adversarial radar signal, we further propose the STDS attack method, which bridges the irreversibility gap in time-frequency analysis. As a result, according to our proposed method, different representations of adversarial radar signals in both the time domain and time-frequency domain can be obtained, greatly enhancing the practicality of the attack.\par

To further demonstrate the effectiveness of our proposed method, we design the DITIMI-FGSM attack method using adversarial attack enhancement strategy. We conducted experiments to test the transferability of the adversarial examples on commonly used lightweight models. Our results indicate that the improved attack method can generate adversarial examples with excellent black-box attack performance on other lightweight models and similar structural models. For instance, we found that the adversarial examples generated by several local surrogate models can achieve over 60$\%$ transfer attack success rates on models such as Mobilenet\_v2 and Efficientnet\_b0.

\section{RELATED WORK}
\subsection{Radar signal classification}
Radar signal is a typical non-stationary signal which we can get corresponding time-frequency image through time-frequency analysis. Benefit from the ability of extracting features automatically from time-frequency images, the radar signal classification method based on DNN has almost eliminated the traditional handcrafted feature extraction methods. Li et al.\cite{c1} combines the attention mechanism with the existing CNN to propose CNN-1D-AM, which can recognize seven kind of radar emitter signals. Zhang et al.\cite{c2} designs adaptive architecture CNN to extract features from CWD of signals. Si et al. \cite{cr4} proposes multiscale lightweight model, which can improve recognition speed on the mobile hardware. In this work, we choose Short-Time Fourier transform (STFT), as our analysis method to obtain time-frequency images. Its processing process is shown in Eq. (1):
\begin{equation}
STFT(t,f) = \int_{ - \infty }^\infty  {x(\tau )h\left( {\tau  - t} \right)} {e^{ - j2\pi ft}}d\tau
\end{equation}
where $h\left(t\right)$  is window function, $x\left(t\right)$ is time-domain radar signal and $\tau$ is fixed length. Note that when we get the STFT result of the original signal, there still need operation of colormap to obtain RGB time-frequency images.

\subsection{Adversarial attacks}
Since Szegedy et al.\cite{FGSM} found that neural networks are easily attacked by adversarial examples, which are perturbed signals added intentionally crafted interference, numerous adversarial attack methods have been proposed in recent years \cite{PGD,CW,DIFGSM}. In general, these attack methods can be categorized into two types, i.e. signal-step attacks\cite{FGSM} and iterative attacks \cite{PGD}. Empirically, iterative attacks can achieve higher success rates in the white-box setting, where the attackers have full knowledge of the structure and parameters of the network. However, if these adversarial examples are tested on a different network, i.e., the black-box setting, single-step attacks perform better. In order to improve the transferability of iterative methods, scholars use momentum \cite{c7}, input transformation \cite{DIFGSM}, ensemble of networks, etc. to reduce the overfitting of the attack algorithm to the local model.\par

\begin{figure}[htbp]
\centerline{\includegraphics[height=5.4cm,width=8.5cm]{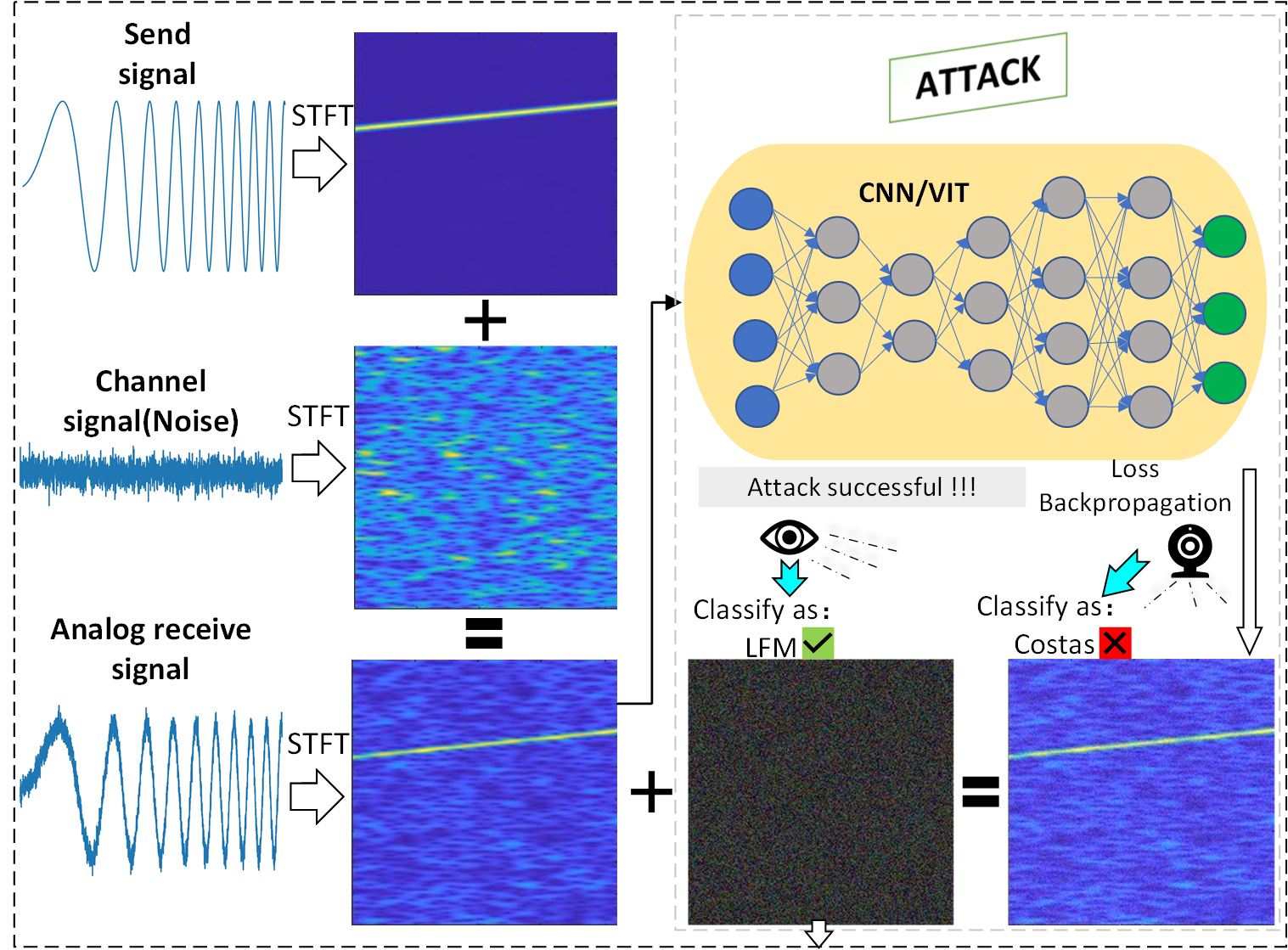}}
\caption{The attack pipeline in the time-frequency image scenario.}
\label{fig1}
\vspace{-0.8em}
\end{figure}

\subsection{Radar signal attack}
Traditional electronic countermeasures (EMC) can be categorized into two types, i.e. suppress interference and deception jamming\cite{c4}. Generally, suppress interference often transmits interference signal with extremely high power, which can make it difficult for enemy radar to capture the echo signal and thus cannot obtain useful signals for recognizing. While deception jamming is aimed to transmit fake signal, it can mislead enemy radar to make wrong decision. It can be
seen that under the premise of ensuring the success of the attack, traditional electronic countermeasures do not constrain perturbed signal.\par

Some works in this field have noticed existing DNN-based methods are vulnerable to adversarial examples, but existing related work operates on signals in the time domain\cite{x1, x2}, taking the received and sampled time domain signal as input to the DNN. They never discuss the feasibility of attacks in the context of time-frequency analysis, which is already the most important step in analyzing radar signals.\par

Different from traditional electronic countermeasures, adversarial attack usually constrains the perturbed signal on the premise of ensuring the success of the attack. Let $I$ be a benign original data, $l$ the corresponding true label and $M\left(I;\theta\right)$ the classifier with parameter $\theta$ that outputs the prediction result. The purpose of the attack is to find adversarial example that is highly similar to the original data but can mislead the classification model under the specified perturbation constraints, as described by Eq. (2):
\begin{equation}
M(I + \rho ) \to \mathop l\limits^ -  {\rm{  }}  s.t.  {\rm{ }}\mathop l\limits^ -   \ne l,{\rm{ }}{\left\| \rho  \right\|_p} < \eta
\end{equation}
Here $\rho$ denotes perturbed data, ${\left\| \rho  \right\|_p} < \eta$ denotes the $\rho-norm$ distance of $\rho$ is subjected to perturbation constraints $\eta$.\par

In radar signal classification task application scenario discussed in this paper, we select cross-entropy as the attack loss function for subsequent generation of adversarial examples, which we denote as $f$.

\section{Methodology}

\begin{figure*}[htbp]
\centerline{\includegraphics[height=5.0cm,width=18cm]{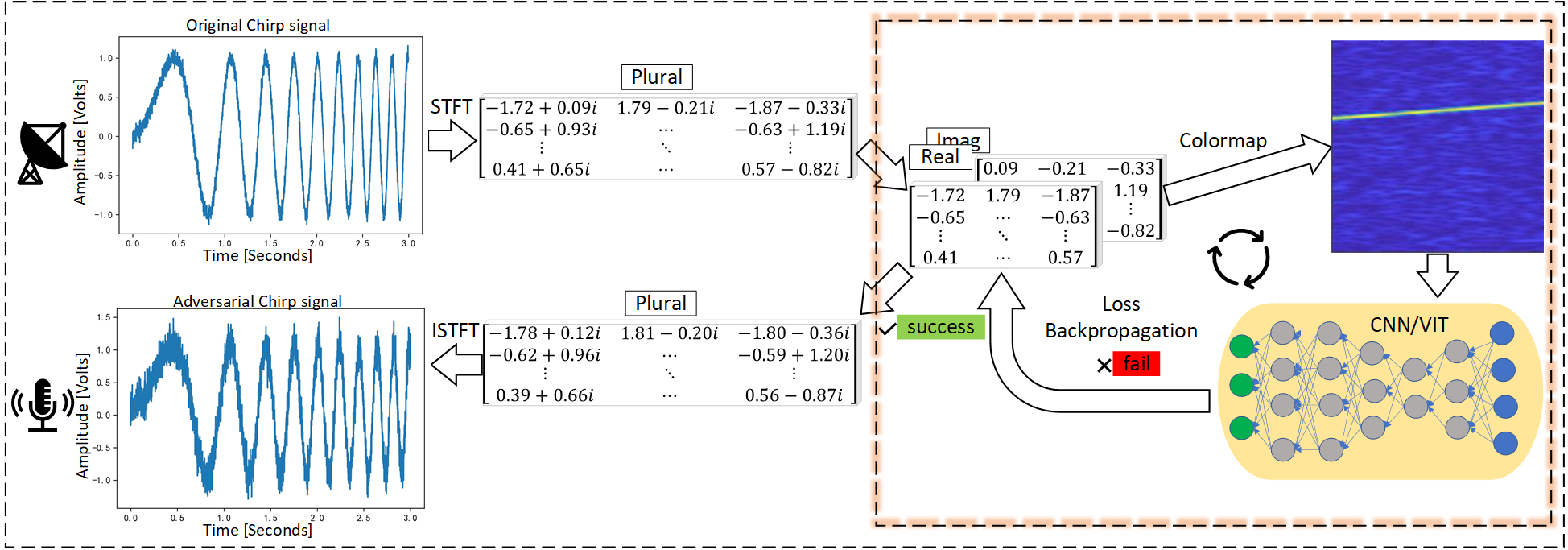}}
\caption{The pipeline of STFT-based time domain signal attack(STDS).}
\label{fig2}
\end{figure*}

\subsection{Image-based time-frequency domain signal attack}
In general signal transmission-reception scenarios, we usually assume transmitter has the information of the transmitted signal and the channel characteristics (noise characteristics) \cite{c9}. The analog received signal can be converted into RGB three-channel time-frequency image through STFT, so we propose the attack pipeline in the time-frequency image scenario. As shown in Fig 1, using the time-frequency image of the analog received signal as input for DNN, through our adversarial attack algorithm, the time-frequency image of the adversarial example that is human-imperceptible but can deceive the model is finally obtained, the 
difference between input and output is the time-frequency image of the perturbed noise signal. In order to obtain better attack effect, we hope adversarial 
examples have strong transferability. We have the following two considerations:

\subsubsection{Momentum-based iterative gradient}
The attack effect of iterative-based\cite{PGD} attacks or optimization-based attacks\cite{CW} always drop sharply in black-box setting because the gradient update direction is unstable so as to fall into the poor local optima. So we use momentum to accumulate velocity vector in the gradient direction in each step.
\subsubsection{Input augmentation}
Another reason for the restriction of adversarial examples’ transferability is iterative attacks tend to overfit current model parameters. Inspired by data augmentation strategy, we apply stochastic image transformations to the inputs with the probability $p$ at each iteration to alleviate overfitting phenomenon. In our experiment, we adopt stochastic image transformations as Eq.(3) and the same implementation details as \cite{DIFGSM} and \cite{TIFGSM}.\par

In summary, we combine momentum-based iterative gradient with input augmentation to obtain the DITIMI-FGSM, it is summarized in Algorithm 1.
\begin{algorithm}[!h]
    \setlength{\abovedisplayskip}{3pt}
    \setlength{\belowdisplayskip}{3pt}
    \caption{DITIMI-FGSM Attack}
    \label{alg:AOS}
    \renewcommand{\algorithmicrequire}{\textbf{Input:}}
    \renewcommand{\algorithmicensure}{\textbf{Output:}}
    \begin{algorithmic}[1]
        \REQUIRE A classifier $M$ with parameters $\theta$ and step size $\alpha$; 
        \REQUIRE Loss function $f$; transformation function ${f_{trans}}$; probability $p$; kernel matrix $W$ and decay factor $\mu$;
        \REQUIRE Time-frequency image $x$ and ground-true label $y$;
        \REQUIRE The size of perturbation $\epsilon$ and iterations $T$. 
        \ENSURE  Time-frequency adversarial noise image ${x_{noise}}$   

        \STATE  ${g^0} = 0; x_{adv}^0 = x$;
        \FOR{$t$ = 0 to $T$-1}
            \STATE $x_{adv}^t = {f_{trans}}(x_{adv}^t;p)$;
            \STATE Calculate the gradient ${\nabla _{x_{adv}^t}}f(x_{adv}^t;y;\theta )$;
            \STATE Update ${g^t}$ by accumulating the velocity vector in the gradient direction as
            \begin{equation}
            {g^{t + 1}} = \mu {g^t} + \frac{{W*{\nabla _{x_{adv}^t}}f(x_{adv}^t;y;\theta )}}{{||W*{\nabla _{x_{adv}^t}}f(x_{adv}^t;y;\theta )|{|_1}}};
            \end{equation}
            \STATE Update $x_{adv}^t$ by applying the sign gradient as
            \begin{equation}
            x_{adv}^{t + 1} = Clip_{x - \varepsilon }^{x + \varepsilon }\{ x_{adv}^t + \alpha sign({g^{t + 1}})\};
            \end{equation}
        \ENDFOR
        \RETURN ${x_{noise}} = x_{adv}^{T - 1} - x$.
    \end{algorithmic}
\end{algorithm}

\begin{figure*}
\centerline{\includegraphics[height=3.5cm,width=18cm]{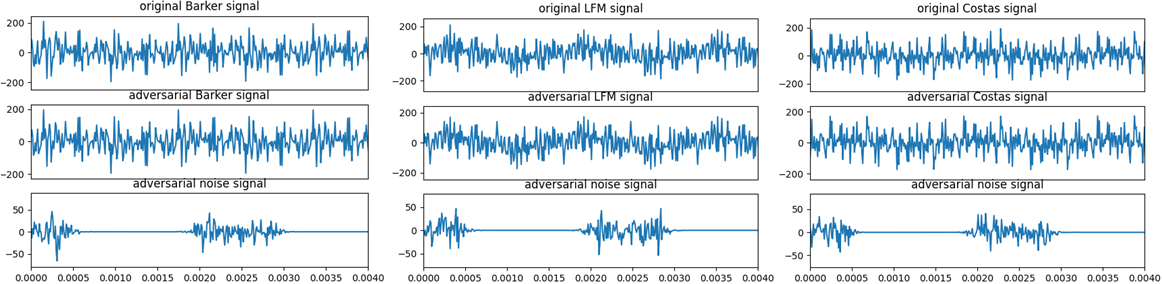}}
\caption{The original time-domain signal, the STDS-generated adversarial time-domain signal, and the corresponding perturbed time-domain signal.}
\end{figure*}

\subsection{STFT-based time domain signal attack}
Now, we obtain the time-frequency image result of the perturbed signal, it is natural to think if we can do inverse transformation (ISTFT) on the time-frequency image, we can get the time-domain perturbed signal. But unfortunately, the inverse transformation from time-frequency image to time domain signal cannot be achieved due to the process of converting the STFT result of the time-domain signal into the time-frequency image only requires modulo operation on plural, while ISTFT requires recover the real and imaginary parts of the plural according to the modulo value, obviously, it is impossible to achieve such an unequal mapping relationship.\par
In order to solve the problem that ISTFT cannot be directly performed, we design time domain attack framework shown in Fig 2. The yellow box is our attack process, both input and output contain all the information required by ISTFT. Therefore, after we obtain the output according to the attack algorithm, we can obtain the time domain form of the perturbed signal through ISTFT. At the beginning of attack, we separate the real and imaginary parts of the STFT result of the analog received signal. Then we use color mapping to convert the split data into RGB three-channel image, which is the time-frequency image we used before. As described in Eq. (5):
\begin{equation}
\begin{aligned}
{\nabla _{inpu{t^t}}}f(x_{adv}^t,y;\theta ) = \frac{{\partial f(x_{adv}^t,y;\theta )}}{{\partial inpu{t^t}}} \\= \frac{{\partial f(x_{adv}^t,y;\theta )}}{{\partial x_{adv}^t}}\frac{{\partial x_{adv}^t}}{{\partial inpu{t^t}}}
\end{aligned}
\end{equation}

Here $f$ is adversarial example generation algorithm, $inpu{t^t}$ is concatenate matrix for the real and imaginary parts of input, $y$ is ground-true label and $\theta$ is a fixed model parameters. We calculate the loss according to the classification results of the model on time-frequency images, and use the gradient of the loss to the input data as basis for modifying input. It should be noted that the reason for using the time-frequency image as the input of the network instead of the concatenate matrix is to be more in line with the real receiving scenario.The algorithm of STDS is summarized in Algorithm 2.

\begin{algorithm}[!h]
    \caption{STDS Attack}
    \label{alg:AOS}
    \renewcommand{\algorithmicrequire}{\textbf{Input:}}
    \renewcommand{\algorithmicensure}{\textbf{Output:}}
    \begin{algorithmic}[1]
        \REQUIRE A classifier $M$ with parameters $\theta$ and learning rate $lr$;  
        \REQUIRE Attack loss function $f$; STFT function ${f_{STFT}}$; ISTFT function ${f_{ISTFT}}$ and colormap function ${f_{Colormap}}$.
        \REQUIRE Original time-domain signal $x(t)$; ground-true label $y$ and iterations $T$
        \ENSURE  Time-domain adversarial noise ${x_{noise}(t)}$   

        \STATE ${X_{STFT}} = {f_{STFT}}(x(t))$;
        \STATE $X_{STFT\_Real}^0={X_{STFT\_Real}}$,\\
               $X_{STFT\_Imag}^0={X_{STFT\_Imag}}$;
        \STATE $Inpu{t^0} = \left[ {\left[ {X_{STFT\_Real}^0} \right],{\rm{ }}\left[      {X_{STFT\_Imag}^0} \right]} \right]$;
        \FOR{$t$ = 0 to $T$-1}
            \STATE $x_{adv}^t = {f_{Colormap}}(Inpu{t^t})$;
            \STATE Calculate the gradient ${{\rm{g}}^t} = {\nabla _{Inpu{t^t}}}f(x_{adv}^t,y;\theta )$;
            \STATE Update ${Inpu{t^t}}$ by gradient descent \\ $Inpu{t^{t + 1}} = Inpu{t^t} + lr{{\rm{g}}^t}$;
        \ENDFOR
        \STATE $Output = \left[ {X_{STFT\_Real}^{{\rm{ }}T - 1} + X_{STFT\_Imag}^{{\rm{ }}T - 1}*i} \right]$;
        \STATE ${x^{'}}(t) = {f_{ISTFT}}(Output)$;
        \RETURN ${x_{noise}}(t) = {x^{'}}(t) - x(t)$.
    \end{algorithmic}
\end{algorithm}

\section{Experiments}
\subsection{Experimental Setup}
\subsubsection{Dataset and models}
We simulate three common radar signals i.e. Barker, Costas, and LFM, all signals’ SNR coverage from -10dB to 10dB. As for models, we select five SOTA models with different parameter quantities and structures as local surrogate models and we include four of the most widely used lightweight models for mobile devices as victim models.
\subsubsection{Attack methods}
We select four different mechanisms attack algorithm i.e. single-step attack FGSM \cite{FGSM}, iterative-based attack PGD\cite{PGD}, optimization-based CW\cite{CW} and our improved algorithm DITIMI-FGSM, which aims to improve the transferability of adversarial examples. We set the same hyper-parameters for each method as \cite{FGSM,PGD,CW,DIFGSM,TIFGSM}.
\subsubsection{Else}
We set $lr$=0.001, step size $\alpha$=1/255, the maximum perturbation $\epsilon$=10.(The change in each pixel value does not exceed 10). We select 'parula' as colormap function ${f_{Colormap}}$.

\subsection{Experimental results}
First, we use different adversarial attack methods (FGSM, PGD, CW, DITIMI-FGSM in time-frequency images reception scenario, and STDS in time domain signal reception scenario) to generate adversarial examples on a local single model with bigger number of parameters, and then test the attack success rate of the adversarial examples on four commonly used lightweight models. This is done to be more in line with actual usage scenarios. The $success$ $rates$, which refers to the probability of adversarial examples being misclassified by victim model are shown in Table 1. The results indicate that almost all attack methods have a good white-box attack success rate, which is close to 100\%. Our DITIMI-FGSM and STDS methods also achieved good attack effects in black-box attacks in two receiving scenarios (with an average attack success rate of over 60\%), indicating that the adversarial examples generated by these two attack methods have high transferability.\par

\begin{figure}[htbp]
\includegraphics[height=4.8cm,width=8.7cm]{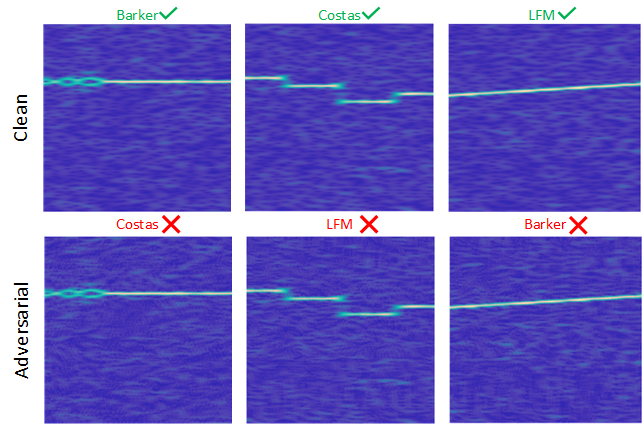}
\caption{Visualization of randomly selected clean images and their corresponding adversarial examples which are generated on Resnet18 using our STDS attack method. All adversarial examples are misclassified by the model, indicating that the attack is successful.}
\label{fig3}
\vspace{-0.8em}
\end{figure}

\begin{table*}[t]
\centering
\caption{The success rates (\%) on four lightweight models by various attacks. The adversarial examples are crafted on Googlenet, Resnet18, ConvNeXt{\_}tiny, Vit{\_}b{\_}16, and VGG11 respectively. Self-attack indicates white-box attack, others indicate black-box attacks. Experiment results demonstrate that adversarial examples have strong transferability.}
\begin{tabular}{@{}cclllll@{}}
\hline
Model & Attack & Self-attack & \begin{tabular}[c]{@{}c@{}}Shufflenet\_v2\cite{shufflenet}\\ 0.34M\end{tabular} & \begin{tabular}[c]{@{}c@{}}Mobilenet\_v2\cite{mobilenet}\\ 2.23M\end{tabular} & \begin{tabular}[c]{@{}c@{}}Efficientnet\_b0\cite{efficientnet}\\ 4.01M\end{tabular} & \begin{tabular}[c]{@{}c@{}}Densenet121\cite{densenet}\\ 6.69M\end{tabular} \\ \hline
\multirow{5}{*}{\begin{tabular}[c]{@{}c@{}}Googlenet\cite{googlenet}\\ 5.60M\end{tabular}} & FGSM & 39.17 & 10.33 & \textbf{66.67} & 66.50 & 32.83 \\
 & PGD & \textbf{100.00} & 23.50 & 56.67 & 60.50 & 9.17 \\
 & CW & \textbf{100.00} & 0.16 & 23.17 & 54.33 & 0.50 \\
 & \textbf{DITIMI-FGSM(Ours)} & \textbf{100.00} & \textbf{67.33} & \textbf{66.67} & 57.33 & \textbf{66.67} \\ \cdashline{2-7}[1pt/3pt] 
 & \textbf{STDS(Ours)} & \textbf{100.00} & 67.00 & 54.00 & \textbf{67.00} & 63.00 \\ \hline
\multirow{5}{*}{\begin{tabular}[c]{@{}c@{}}Resnet18\cite{resnet}\\ 11.18M\end{tabular}} & FGSM & 63.00 & 44.83 & \textbf{66.33} & 63.83 & 35.33 \\
 & PGD & \textbf{100.00} & 53.83 & 35.00 & 66.67 & 0.16 \\
 & CW & \textbf{100.00} & 12.67 & 12.17 & 39.83 & 0.33 \\
 & \textbf{DITIMI-FGSM(Ours)} & \textbf{100.00} & \textbf{66.67} & 46.83 & \textbf{84.00} & \textbf{66.67} \\ \cdashline{2-7}[1pt/3pt] 
 & \textbf{STDS(Ours)} & 80.00 & 53.00 & 47.00 & 32.00 & 63.00 \\ \hline
\multirow{5}{*}{\begin{tabular}[c]{@{}c@{}}ConvNeXt\_tiny\cite{convnext}\\ 27.82M\end{tabular}} & FGSM & 5.00 & 0.33 & 56.67 & 66.17 & 13.00 \\
 & PGD & 94.83 & 2.83 & 56.67 & 66.33 & 6.17 \\
 & CW & \textbf{100.00} & 7.83 & 44.17 & 66.67 & 0.00 \\
 & \textbf{DITIMI-FGSM(Ours)} & \textbf{100.00} & \textbf{66.67} & 65.50 & \textbf{67.17} & 66.00 \\ \cdashline{2-7}[1pt/3pt] 
 & \textbf{STDS(Ours)} & 90.00 & 64.00 & \textbf{67.00} & 57.00 & \textbf{67.00} \\ \hline
\multirow{5}{*}{\begin{tabular}[c]{@{}c@{}}ViT\_b\_16\cite{vit}\\ 85.50M\end{tabular}} & FGSM & 51.83 & 3.67 & \textbf{60.83} & 64.17 & 13.00 \\
 & PGD & \textbf{100.00} & 0.67 & 41.67 & 65.00 & 6.17 \\
 & CW & \textbf{100.00} & 0.00 & 0.33 & 55.63 & 0.00 \\
 & \textbf{DITIMI-FGSM(Ours)} & \textbf{100.00} & \textbf{57.16} & 47.16 & \textbf{69.17} & 66.00 \\ \cdashline{2-7}[1pt/3pt] 
 & \textbf{STDS(Ours)} & \textbf{100.00} & 20.00 & 54.00 & 67.00 & \textbf{67.00} \\ \hline
\multirow{5}{*}{\begin{tabular}[c]{@{}c@{}}VGG11\cite{vgg}\\ 114.08M\end{tabular}} & FGSM & 50.83 & 6.67 & 66.67 & 66.17 & 32.33 \\
 & PGD & \textbf{100.00} & 3.50 & 65.33 & 66.67 & 15.00 \\
 & CW & \textbf{100.00} & 0.33 & 44.17 & 59.83 & 6.67 \\
 & \textbf{DITIMI-FGSM(Ours)} & \textbf{100.00} & \textbf{66.63} & 39.00 & \textbf{67.00} & 66.67 \\ \cdashline{2-7}[1pt/3pt] 
 & \textbf{STDS(Ours)} & \textbf{100.00} & 65.00 & \textbf{67.00} & \textbf{67.00} & \textbf{67.00} \\ \hline
\end{tabular}
\end{table*}

We compare time-frequency image of the adversarial signal with the time-frequency image of the corresponding original time domain signal(results are shown in Fig. 4) to show the result is human-imperceptible. Since the colormap process is non-linear, the corresponding relationship between the pixel change of the time-frequency image and the gradient information loses its physical meaning, so here we use the $\ell_2$ norm to measure average perturbation. The $\ell_2$ of the average perturbation is 2.286, which shows that the perturbation is concealed. Visually, the perturbations are only manifested as some noise points on the time-frequency images, without affecting the quality of the image display.\par

In Fig. 3, we present some of the original time domain signals used in the experiment, the time domain adversarial example signals generated by the STDS attack method, and the difference between them, namely the perturbed time domain signal. It can be seen that there is little difference between the time domain signals before and after the change, and the perturbed signal is a non-persistent signal (which is advantageous for generating through simulation means). This indicates that our attack method has high practical usability. It is important to compare the differences between the time-frequency images before and after the signal changes in the time domain signal reception scenario. This is because in this scenario, the received time domain signal will undergo time-frequency analysis, and the receptor essentially still recognizes the signal's time-frequency images. Therefore, ensuring that the differences in the time-frequency image changes are small can avoid the receptor from becoming suspicious of potential attacks.\par

When using DNNs to develop radar signal recognition applications, most developers choose to use existing model structures for development and fine-tuning, which leads to some similarities among victim models. We select different models from the Resnet and Vgg series, use different attack methods to generate adversarial examples, and then test the transferability of adversarial examples on models with similar structures. The success rate of the attack is shown in Table 2. As the results indicate, adversarial examples generated by FGSM, DITIMI-FGSM, and STDS exhibit high transferability across models with similar structures. Notably, our proposed DITIMI-FGSM and STDS methods can achieve a black-box attack success rate of over 50\% in most cases while ensuring a 100\% white-box attack success rate. For example, the adversarial examples generated by the DITIMI-FGSM method on Resnet152 can achieve attack success rates of 55.33\%, 66.67\%, and 66.83\% on Resnet18, Resnet50, and Resnet101, respectively. \par

\begin{table*}[t]
\centering
\caption{The success rates (\%) on models with similar structures by FGSM, DITIMI-FGSM and STDS attack. Experiment results demonstrate that similar model structure can provide more possibilities for transferability.}
\begin{tabular}{@{}c|ccccc|ccccc@{}}
\hline
Attack & Model & \begin{tabular}[c]{@{}c@{}}Resnet18\\ 11.18M\end{tabular} & \begin{tabular}[c]{@{}c@{}}Resnet50\\ 23.51M\end{tabular} & \begin{tabular}[c]{@{}c@{}}Resnet101\\ 42.15M\end{tabular} & \begin{tabular}[c]{@{}c@{}}Resnet152\\ 58.15M\end{tabular} & Model & \begin{tabular}[c]{@{}c@{}}VGG11\\ 114.08M\end{tabular} & \begin{tabular}[c]{@{}c@{}}VGG13\\ 114.27M\end{tabular} & \begin{tabular}[c]{@{}c@{}}VGG16\\ 119.58M\end{tabular} & \begin{tabular}[c]{@{}c@{}}VGG19\\ 124.89M\end{tabular} \\ \hline
FGSM & \multirow{3}{*}{\begin{tabular}[c]{@{}c@{}}Resnet18\\ 11.18M\end{tabular}} & 63.17 & 41.83 & 12.17 & 55.17 & \multirow{3}{*}{\begin{tabular}[c]{@{}c@{}}VGG11\\ 114.08M\end{tabular}} & 45.83 & 30.16 & 5.00 & 1.70 \\
\textbf{DITIMI-FGSM(Ours)} &  & \textbf{100.00} & 66.67 & \textbf{56.33} & \textbf{63.67} &  & \textbf{100.00} & \textbf{96.83} & \textbf{57.50} & \textbf{63.67} \\
\textbf{STDS(Ours)} &  & \textbf{100.00} & \textbf{67.00} & 52.00 & 35.00 &  & \textbf{100.00} & 82.00 & 44.00 & 60.00 \\ \hline
FGSM & \multirow{3}{*}{\begin{tabular}[c]{@{}c@{}}Resnet50\\ 23.51M\end{tabular}} & 46.67 & 65.00 & 28.33 & 60.33 & \multirow{3}{*}{\begin{tabular}[c]{@{}c@{}}VGG13\\ 114.27M\end{tabular}} & 18.30 & 41.67 & 1.70 & 3.30 \\
\textbf{DITIMI-FGSM(Ours)} &  & \textbf{66.63} & \textbf{100.00} & \textbf{66.00} & \textbf{66.50} &  & \textbf{91.17} & \textbf{100.00} & \textbf{50.00} & 35.33 \\
\textbf{STDS(Ours)} &  & 42.00 & \textbf{100.00} & 55.00 & 30.00 &  & 75.00 & \textbf{100.00} & 37.00 & \textbf{52.00} \\ \hline
FGSM & \multirow{3}{*}{\begin{tabular}[c]{@{}c@{}}Resnet101\\ 42.15M\end{tabular}} & 40.17 & 55.83 & 48.50 & 62.67 & \multirow{3}{*}{\begin{tabular}[c]{@{}c@{}}VGG16\\ 119.58M\end{tabular}} & 17.00 & 5.00 & 15.67 & 0.00 \\
\textbf{DITIMI-FGSM(Ours)} &  & \textbf{58.88} & \textbf{66.67} & \textbf{100.00} & \textbf{70.50} &  & 53.50 & \textbf{92.00} & \textbf{100.00} & \textbf{64.17} \\
\textbf{STDS(Ours)} &  & 49.00 & 32.00 & \textbf{100.00} & 12.00 &  & \textbf{54.00} & 52.00 & \textbf{100.00} & 62.00 \\ \hline
FGSM & \multicolumn{1}{l}{\multirow{3}{*}{\begin{tabular}[c]{@{}l@{}}Resnet152\\ 58.15M\end{tabular}}} & 36.33 & 60.67 & 32.83 & 66.17 & \multicolumn{1}{l}{\multirow{3}{*}{\begin{tabular}[c]{@{}l@{}}VGG19\\ 124.89M\end{tabular}}} & 10.00 & 2.80 & 3.70 & 5.00 \\
\textbf{DITIMI-FGSM(Ours)} & \multicolumn{1}{l}{} & 55.33 & \textbf{66.67} & \textbf{66.83} & \textbf{100.00} & \multicolumn{1}{l}{} & \textbf{16.17} & \textbf{41.17} & \textbf{33.83} & \textbf{100.00} \\
\textbf{STDS(Ours)} & \multicolumn{1}{l}{} & \textbf{100.00} & 57.00 & 27.00 & \textbf{100.00} & \multicolumn{1}{l}{} & 10.00 & 8.00 & 17.00 & \textbf{100.00} \\ \hline
\end{tabular}
\end{table*}

\section{Conclusions}
In this paper, we propose a new paradigm of electronic countermeasures for radar signals using adversarial attacks in the context of DNN-based radar signal classification. We propose the DITIMI-FGSM method to enhance the transferability of adversarial examples, thereby achieving better attack performance. To obtain the time domain representation of the adversarial interference signal in the time-frequency analysis scenario, we propose the STDS method. Experimental results show that the adversarial examples generated by our proposed method minimally modify the original signal and are highly transferable, indicating their practicality.\par
There are mainly two challenges to be solved in the future: 1.New attack method which can further limit perturbation in time domain should be investigated. 2.The experiment in this
paper is conducted on simulated data. Hence the generalization capability of our attack need to be investigated.

\section*{Acknowledgment}
This work was supported in part by Beijing Natural Science Foundation (L191004) and the National Natural Science Foundation of China (Grant no. 62171025).

\bibliographystyle{IEEEtran}
\bibliography{ref.bib}

\end{document}